\documentclass[aps,pra,twocolumn,showkeys,amsmath,amssymb]{revtex4}
\usepackage{graphicx,amsmath,amsfonts,amssymb,float,bm,epsfig,color}

\newcommand{\newc}{\newcommand}

\newc{\be}{\begin{equation}}
\newc{\ee}{\end{equation}}
\newc{\bea}{\begin{eqnarray}}
\newc{\eea}{\end{eqnarray}}
\newc{\beas}{\begin{eqnarray*}}
\newc{\eeas}{\end{eqnarray*}}

\newc{\pardt}{\partial_{t}}
\newc{\pardxi}{\partial_{i}}
\newc{\pardts}{\partial_{t^{*}}}
\newc{\pardxis}{\partial_{i^{*}}}
\newc{\pardxj}{\partial_{j}}
\newc{\pardxk}{\partial_{k}}
\newc{\pard}{\partial}
\newc{\ti}{\tilde}

\newc{\s }{\overline}

\newc{\sect}{\section}
\newc{\subs}{\subsection}

\newc{\defi}{\definition}
\newc{\prop}{\proposition}
\newc{\rem}{\remark}
\newc{\lem}{\lemma}
\newc{\exa}{\example}
\newc{\theo}{\theorem}
\newc{\coro}{\corollary}
\newc{\post}{\postulate}
\newc{\state}{\statement}
\newc{\ds}{\displaystyle}

\def\j{{\textrm{i}}}

\newbox\tempbox

\begin{document}
\baselineskip0.5cm
\renewcommand {\theequation}{\thesection.\arabic{equation}}

\title{Matter-wave dark solitons in box-like traps}

\author{M.~Sciacca\footnote{Corresponding author.}}
    \email{michele.sciacca@unipa.it}
    \affiliation{Dipartimento Scienze Agrarie e Forestali, Universit\`a di Palermo, Viale delle Scienze, 90128 Palermo, Italy}
    \affiliation{Istituto Nazionale di Alta Matematica, Roma 00185, Italy}
    \author{C. F. Barenghi}
    \affiliation{Joint Quantum Centre Durham-Newcastle, School of Mathematics and Statistics, Newcastle University, Newcastle upon Tyne, NE1 7RU, United Kingdom}
    \author{N. G. Parker}
     \affiliation{Joint Quantum Centre Durham-Newcastle, School of Mathematics and Statistics, Newcastle University, Newcastle upon Tyne, NE1 7RU, United Kingdom}


\date{}

\begin{abstract}
Motivated by the experimental development of quasi-homogeneous 
Bose-Einstein condensates confined in box-like traps,
we study numerically the dynamics of dark solitons in
such traps at zero temperature.  We consider the cases where the side walls of the box potential rise either as a power-law or a Gaussian.  While the soliton propagates through the homogeneous interior of the box without dissipation, it typically dissipates energy during a reflection from a wall through the emission of sound waves, causing a slight increase in the soliton's speed.  We characterise this energy loss as a function of the wall parameters.  Moreover, over multiple oscillations and reflections in the box-like trap, the energy loss and speed increase of the soliton can be significant, although the decay eventually becomes stabilized when the soliton equilibrates with the 
ambient sound field.
\end{abstract}

\keywords{dark soliton; Bose-Einstein condensate; soliton-sound interaction; soliton decay; box-like trap.}

\maketitle

\section{Introduction}\label{sec: Intro}
Dark solitons are one-dimensional non-dispersive waves which arise in 
defocussing nonlinear systems as localized depletions of the field 
envelope \cite{Kivshar1998}.   To date, they have been observed in 
systems ranging from
optical fibres \cite{Emplit,Krokel,Weiner}, magnetic films \cite{Chen}, 
plasmas \cite{Heidemann}, waveguide arrays \cite{Smirnov}, 
to water \cite{Chabchoub}  and atomic Bose-Einstein 
condensates \cite{Frantzeskakis2010}.  This work is concerned with the 
last system; here the matter field of the gas experiences a 
defocussing cubic nonlinearity arising from the repulsive short-range 
atomic interactions.  In the limit of zero temperature, the  mean 
matter field is governed by a cubic nonlinear Schr\"dinger equation 
(NLSE) called the Gross-Pitaevskii equation (GPE) 
\cite{Stenholm-PR363-2002heuristic, Carretero-N21-2008nonlinear, 
Gross-NC20-1961structure, Gross-1963, Pitaevskii-1961}.  
Many experiments have generated and probed these matter-wave 
dark solitons \cite{Burger1999,Denschlag2000,Dutton2001,Engels2007,Jo2007,Becker2008,Chang2008,Stellmer2008,Weller2008,Aycock2016}.  

A necessary feature of an atomic condensate is  the
trapping potential required to confine  it in space.   
When the trapping potential is highly elongated  in one direction compared
to the other two, the condensate  becomes effectively one-dimensional, 
and its longitudinal dynamics  is described by the 1D GPE.  
If the system is homogeneous in the longitudinal direction, the 
GPE is integrable and supports exact dark soliton solutions.  
Dark solitons appear as a  local notch in the atomic 
density  with a phase slip  across it, and travel with 
constant speed \cite{Reinhardt1997,Frantzeskakis2010}  while retaining
their shape.  However, the presence  of confinement in the
longitudinal direction  breaks the ``complete integrability'' of the governing
equation and 
causes the dark soliton to decay  via the emission of 
sound waves \cite{Busch2000,Huang2002,Parker2003,Pelinovsky2005,Parker2010}.
An analogous effect arises in nonlinear optics due to inhomogeneities 
of the optical nonlinearity \cite{Kivshar1998,Pelinovsky1996}.  
In condensates, dark solitons may also decay  through thermal 
dissipation \cite{Fedichev1999,Muryshev2002,Jackson2007} and 
transverse `snaking' instability into vortex pairs or rings 
\cite{Muryshev1999,Carr2000,Feder2000,Anderson2001,Dutton2001,Theocharis2007,Hoefer2016}; 
both decay channels can be effectively eliminated 
by operating at ultracold temperatures and in tight 1D geometries, 
respectively.

To date, the trapping potentials most commonly  used
have been harmonic (quadratic in the distance from the centre 
of the condensate).   Evolution and stability of dark solitons 
moving under a longitudinal harmonic potential  have
been carefully analyzed.   We know that the
soliton tends to oscillate back and forth through the condensate 
at a fixed proportion of the trap's frequency \cite{Fedichev1999,
Muryshev1999,Busch2000,Huang2002,Theocharis2007}.  While the inhomogeneous potential 
leads to sound  emission from the soliton, the harmonic trap uniquely 
supports an equilibrium between sound emission and reabsorption, 
such that the soliton decay is stabilized \cite{Parker2003,Parker2010}. Theoretical work has also considered the 
radiative behaviour of a dark soliton  moving under the effect
of linear potentials and steps \cite{Parker2003b}, 
perturbed harmonic traps \cite{Busch2000,Parker2003,Allen2011}, 
optical lattices \cite{Kevrekidis2003,Parker2004}, 
localized obstacles \cite{Frantzeskakis2002,Parker2003b,Radouani2003,
Radouani2004,Proukakis2004,Bilas2005a,Hans2015}, anharmonic traps 
\cite{Radouani2003,Parker2010} and disordered potentials \cite{Bilas2005b}. 
For slowly-varying potentials,  {it is found that}
the power emitted  by the solition is proportional to the 
square of the soliton's acceleration
\cite{Kivshar1998,Pelinovsky1996,Parker2003}.  

Increasingly, however, experiments are employing box-like traps to 
produce quasi-homogeneous condensates.  Such traps have been realized 
in one \cite{Meyrath,VanEs}, two \cite{Chomaz} and three \cite{Gaunt} 
dimensions (with tight harmonic trapping in the remaining directions 
in the  {1D and 2D cases}). These  new traps
feature flat-bottomed  central regions
and end-cap  potential  provided by optical or 
electromagnetic fields;  the boundaries are therefore
soft, unlike infinite hard walls of existing mathematical models.
For example, the 1D optical box trap of Ref. \cite{Meyrath} featured 
approximately Gaussian walls, while the 2D and 3D optical box traps of 
Refs. \cite{Gaunt,Chomaz} had a power-law scaling in the 
range from $x^{10}$ to $x^{15}$.   
In the bulk of the box-trap, 
where the density is homogeneous, a dark soliton  is expected to 
propagate at constant speed and retain its shape;
however, the 
nature of the reflection of the soliton from  boundaries which are
steeper than the traditional quadratic dependence and softer than
hard boundaries is  still unexplored. Here we seek to address this 
problem through a systematic computational study of the reflection 
of a dark soliton from power-law and Gaussian walls.

\section{Mathematical model}

We consider an atomic condensate in the limit of zero-temperature, with 
arbitrary trapping $V(x)$ along the axis and tight harmonic trapping 
in the transverse directions.  Assuming 
the quasi-1D configuration, the condensate is described
by the one-dimensional wavefunction, $\Psi (x,t)$; the atomic density 
follows as $n( x,t)=|\Psi (x,t)|^2$.  The dynamical evolution equation 
of $\Psi$ is governed by the one-dimensional Gross-Pitaevskii equation,
\be\label{eq:GP1}
	i \hbar \Psi_t= -\frac{\hbar^2}{2m} \Psi_{xx} +V(x)\Psi + g |\Psi|^2 \Psi,
\ee
where $m$ is the atomic mass and the nonlinear coefficient 
$g=4 \pi \hbar^2 a_s/m$ arises from short-range atomic interactions 
of {\it s}-wave scattering length $a_s$, and subscripts denote partial 
derivatives.  

Since we are concerned with quasi-homogeneous condensates,
 it is natural to adopt units relating to the bulk of the 1D condensate, 
where the density is $n_0=\sqrt{\mu_0/g}$ \cite{Primer}, and the
chemical potential $\mu_0$ is the characteristic energy scale.  
The healing length $\xi_0=\hbar/\sqrt{m n_0 g}$ is the minimum spatial 
scale of density variations, and the speed of sound $c = \sqrt{n_0 g/m}$ 
is the typical speed scale; the natural timescale of the bulk condensate 
follows as $\xi_0/c_0$.  Employing these quantities as units leads 
to the following dimensionless GPE \cite{Primer},
\be\label{eq:GP}
	\j u_t=-\frac{1}{2} u_{xx}+|u|^2u +V(x) u,
\ee
where all variables are in their dimensionless form.  
Throughout the rest of this paper we employ dimensionless variables.  

The total energy of the condensate, given by the integral,
\be\label{eq:energy}
	E_{\rm tot}=\int \left(\frac{1}{2}\left|\psi_x \right|^2 + V(x) |\psi|^2+\frac{1}{2}|\psi|^4 \right)~{\rm d}x
\ee
is conserved within the GPE, as confirmed by our numerical simulations.

Equation\;\eqref{eq:GP} have been investigated over the years  
in terms of the  ``complete integrability''  
(see   \cite{BruguarinoJMP51-2010} and reference therein).
This property (even though still not univocally defined) 
regards the existence  of infinite number of  conservation laws 
and the possibility of relating the nonlinear PDE (partial differential 
equation) to a linear PDE by an explicit transformation.
The main  feature is that   Eq. \eqref{eq:GP} is not completely 
integrable 
except for the case  $V(x)=ax+b$, with $a$ and $b$ 
constants \cite{BruguarinoJMP51-2010}.  For $V(x)=0$,  
the GP equation\;\eqref{eq:GP} has the following exact dark 
soliton solution \cite{Kivshar1998,Frantzeskakis2010},
\be\label{sol:GP}
	\ds u= \left\{k \tanh \left[k \left(x-tv -s_0\right)\right]+\j v \right\}e^{-\j \left(t-\theta _0\right)},
\ee
where $k=\sqrt{1-v^2}$ is the amplitude of the dark soliton, $v$ is 
the soliton's speed (and $|v|<1$) and $s_0$ and  $\theta_0$ 
are arbitrary reference values
of the position and phase of the soliton.   
The normalized  dark soliton's energy is \cite{Kivshar1998},
\begin{equation}
	E_{\rm sol}=\frac{4}{3}(1-v^2)^{3/2}.
\label{eq:energy}
\end{equation}

In the absence of the external potential ($V(x)=0$), 
the soliton \eqref{sol:GP} propagates without any loss along the BEC.  
This lossless motion results from the perfect balance of
nonlinear ($|u|^2u$) and linear ($u_{xx}$) terms in Eq. \eqref{eq:GP}.
In optics the soliton is the envelope of
different plane waves of different frequencies and phase velocities
which moves with the group speed $v$, and the two terms induce
self-phase modulation (SPM) and  
group velocity dispersion (GVD), respectively.
When the balance between the two terms ceases or is altered,
some harmonic components acquire more energy or new harmonics are 
generated by the nonlinearity, and what one sees is the 
generation of small-amplitude density (sound) waves (as also explained in Section\;\ref{sec:discussion}).

Our work is based on numerical simulations of the dimensionless 
1D GPE \eqref{eq:GP}.  Numerical time integration of the equation 
is performed using the split-step Fourier method.  
The initial condition consists of the ground state condensate solution 
obtained via the technique of imaginary-time propagation of the GPE, 
into which a dark soliton solution of \eqref{sol:GP} is 
multiplied at the origin 
(this solution is appropriate because at the origin the system is locally
homogeneous).  During the course of the longest simulations (e.g. Fig. \ref{fig:figure6}) the total energy of the system, $E_{\rm tot}$,
changes by less than 1 part in $10^4$.
 
\begin{figure}[htbp]
      \centering
       \includegraphics[width=1\columnwidth]{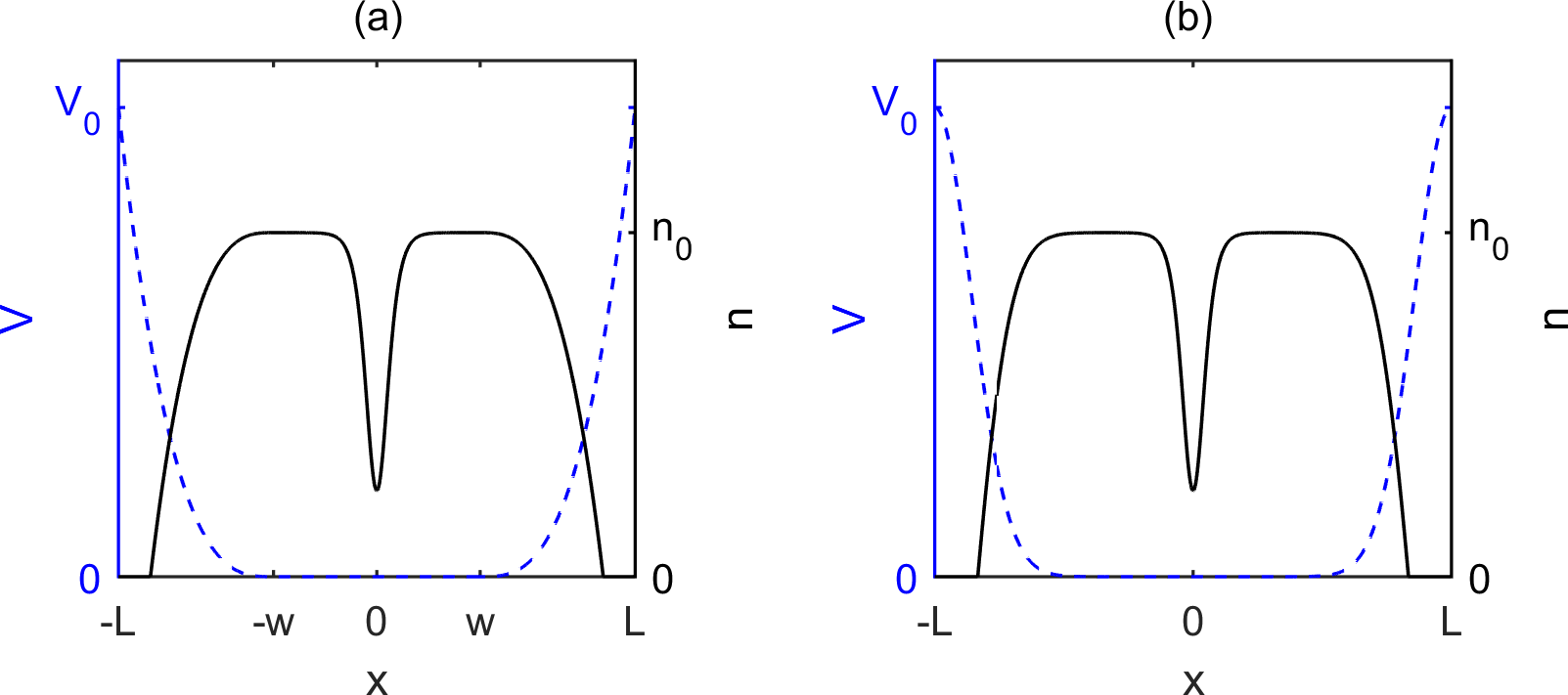}
    \caption{(Color online) Schematic representations
of the potential $V(x)$ and density $n(x)$ in our box traps, 
with a dark soliton at the origin.  
(a) In the {\it power-law box}, the potential is flat ($V=0$) 
over the region $[-w,w]$, and increases as a power-law outside this region, 
reaching the maximum value $V_0=V(L)$ at the edges $x=\pm L$ 
of the box.  
(b) In the {\it exponential box}, Gaussian potentials 
(centred at $x=\pm L$ and with width $L-w$ form 
the end-caps of the box.}
\label{fig:figure1}        
\end{figure}

We consider two types of quasi-homogeneous box potentials.  The first, 
termed the {\it power-law box} and 
motivated by the experiments of Refs. \cite{Gaunt,Chomaz},
is characterised by boundaries where 
the potential increases as a power of the spatial coordinate.
The overall potential has the form,
\be\label{eq:potential}
	V(x)= \left\{ \begin{array}{ll}
 0 &\mbox{ if $|x|\leq w$} \\
  \left(\frac{|x|-w}{l}\right)^\alpha &\mbox{ if $w<|x|\leq L$}
       \end{array} \right.
\ee
where $\alpha$ is the exponent of the potential at the boundaries, 
$2w$ is the width of the flat part
of the potential, and $2L$ is the whole width of the potential.  
This potential is shown schematically in Fig. \ref{fig:figure1}(a).  
The width of the boundary is then $\ds L-w$. The height of the potential 
wall is given by $\ds V_0=V(L)=\left(\frac{L-w}{l}\right)^\alpha$, 
and $l$ is a parameter used to enstablish the height of the side of the 
potential.  The parameters we modify in our numerical experiments are 
the exponent $\alpha$, the height of the boundary potential $V_0$ 
and the width of the boundary potential $\ds L-w$.

The second form of quasi-homogeneous box trap is that where the end-caps 
are formed by laser-induced Gaussian potentials, 
as used in Refs. \cite{Meyrath}.  This box, termed the 
{\it exponential box} and shown schematically 
in Fig. \ref{fig:figure1}(b), has the form,
\be\label{eq:potential-Gauss}
V(x)=V_0\left[{\rm e}^{-\frac{\left(x-L\right)^2}{c^2}} +{\rm e}^{-\frac{\left(x+L\right)^2}{c^2}} \right].
\ee
The crest of the Gaussian potentials are located at 
$x=\pm L$, $V_0$ is their amplitude, and $c$ characterises their 
width.  As in the power-law box, we perform numerical experiments 
to explore the dependence on the amplitude $V_0$ and the width $c$ 
of the boundary potentials on the soliton's motion.

\section{Results}\label{sec:results}

In Section \ref{sec:single} we shall examine a single reflection of a 
dark soliton with a boundary of the box, for both the power-law 
and exponential box types.  Later, in Section\;\ref{sec:multiple}, 
we shall extend our analysis to multiple oscillations and reflections 
in the box.  Throughout this section we set the box width to the 
arbitrary value $L=80$.  

\subsection{Single reflection}\label{sec:single}

\begin{figure}[h]
  \centering
       \includegraphics[width=0.9\columnwidth]{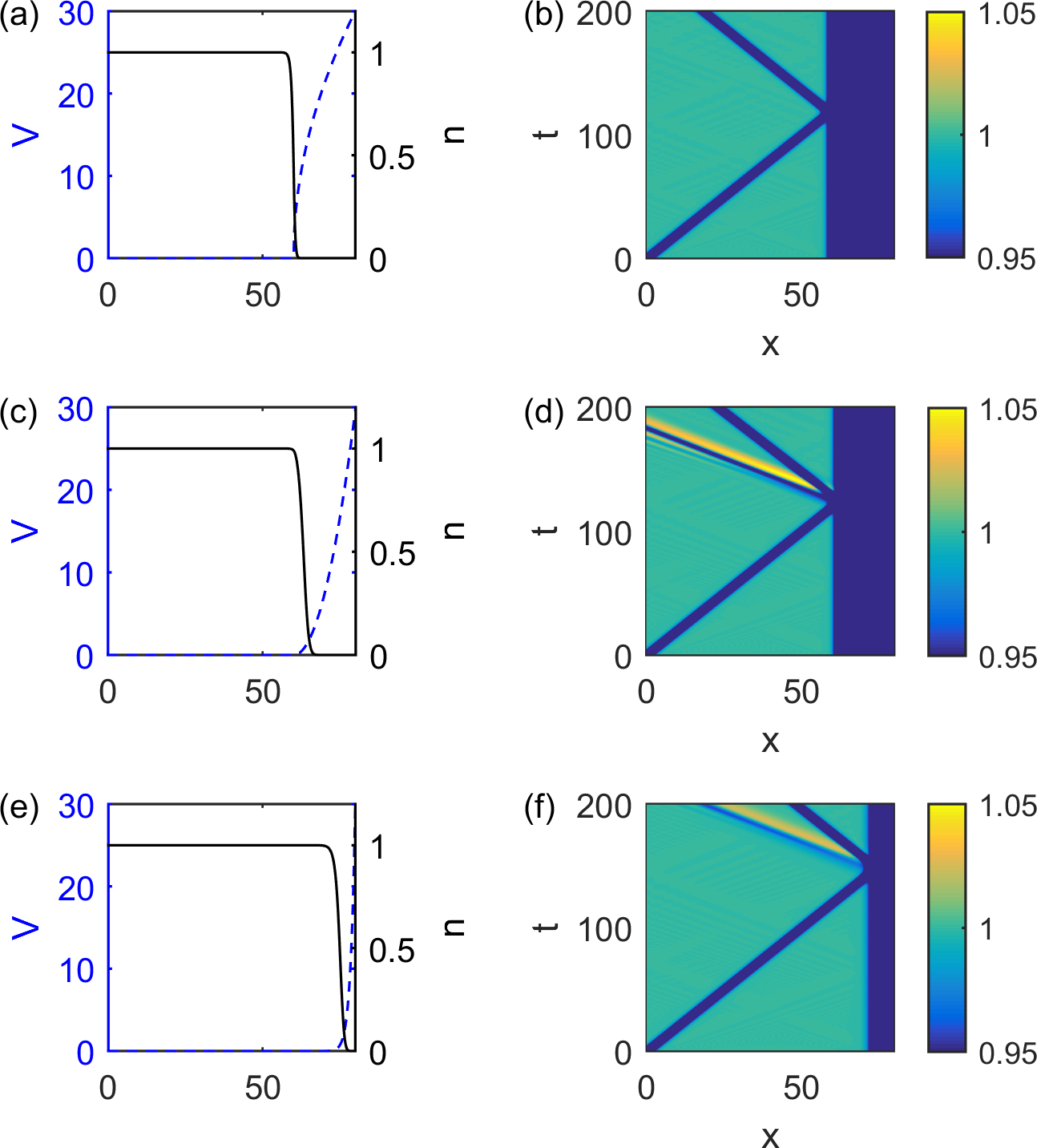}
\caption{(Color online). Examples of the reflection of a $v=0.5$ 
dark soliton from the boundary of a power-law box. 
Shown is (a, c, e) the box potential \eqref{eq:potential} and ground-state 
density profile, and (b, d, f) the evolution of the density during 
the reflection.  Plots (a, b) correspond to a power-law exponent 
of $\alpha=0.5$, (c, d) correspond to $\alpha=2$ and (e, f) 
correspond to $\alpha=13$.  
The remaining parameters are fixed to $L=80$, $w=60$ and $V_0=30$. \label{fig:figure2}}         
\end{figure}

A dark soliton\;\eqref{sol:GP} is introduced at the origin with arbitrary 
speed $v=0.5$ (in the positive $x$ direction) and launched at the $x=L$ 
boundary of the power-law box. Figure\;\ref{fig:figure2} shows the 
dynamics during the reflection from a power-law boundary with fixed 
width $w=60$ and amplitude $V_0=30$, and three different exponents.  
For $\alpha=0.5$ (a, b), the soliton reflects elastically.  
Here the boundary looks effectively like a hard wall - up to the 
typical energy scale of the condensate, $V \sim 1$, 
the potential remains very steep.  For a quadratic potential 
$\alpha=2$ (c, d), however, a pulse of sound waves is generated 
during the reflection which propagate in the negative $x$ direction 
at the speed of sound.  These waves have amplitude of around $\sim 5\%$ 
of the peak density.  For a much higher exponent $\alpha=13$ (e, f), 
again sound waves are emitted during the reflection, 
with a slightly reduced amplitude.  

  \begin{figure}[h]
  \centering
       \includegraphics[width=0.9\columnwidth]{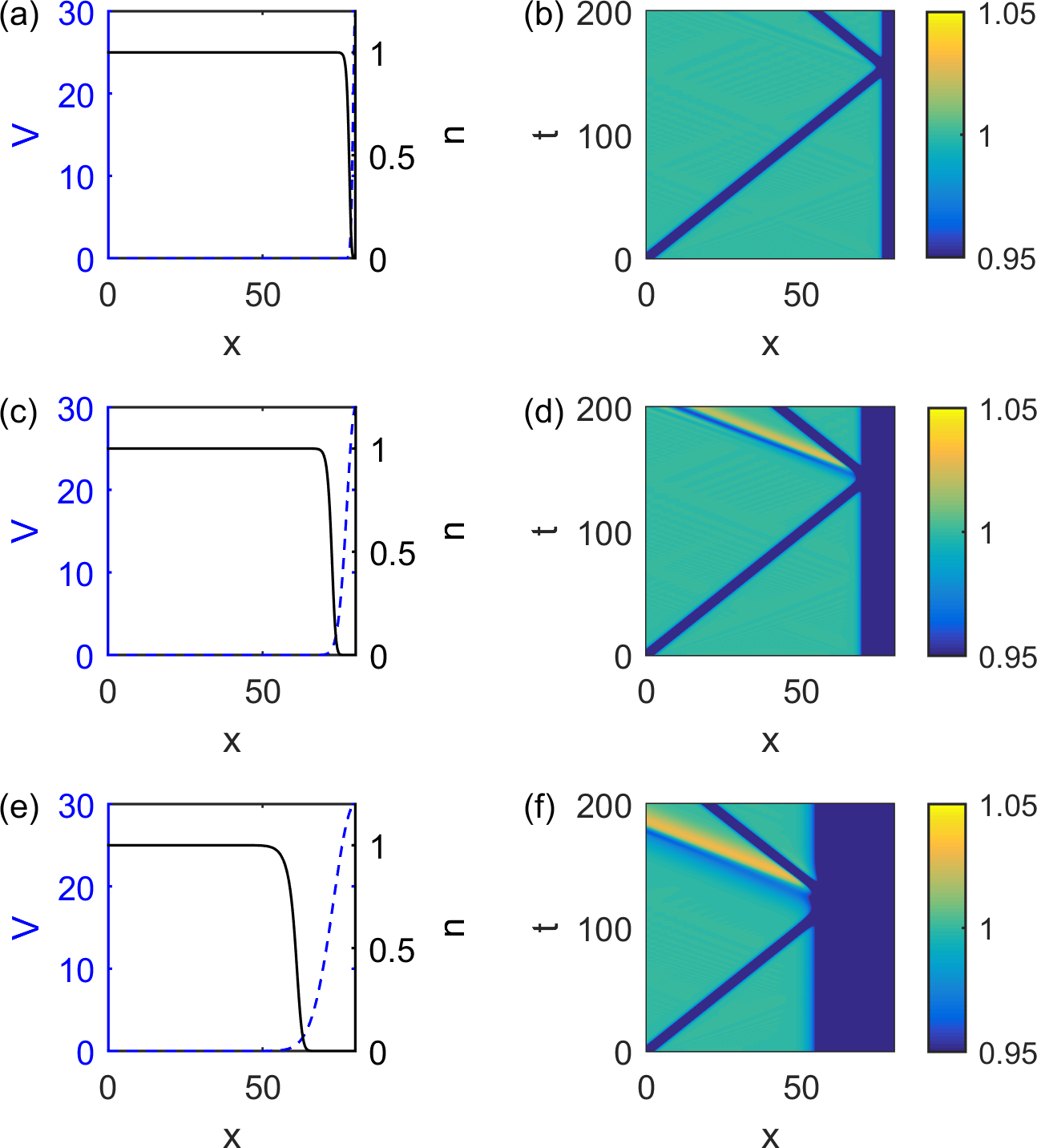}
\caption{(Color online). Examples of the reflection of a $v=0.5$ dark 
soliton from the boundary of an exponential box.  
Shown is (a, c, e) the box potential \eqref{eq:potential} and 
ground-state density profile, and (b, d, f) the evolution of 
the density during the reflection.  Plots (a, b) correspond to a 
Gaussian width of $c=1$, (c, d) correspond to $c=4$ and (e, f) 
correspond to $c=10$.  The remaining parameters are 
fixed to $L=80$ and $V_0=30$. \label{fig:figure3}}         
\end{figure}

Figure \ref{fig:figure3} shows example dynamics for the soliton 
reflecting from an exponential boundary.  
Here the amplitude of the boundaries is fixed to $V(L)=30$ 
and the width $c$ varied.  For a narrow edge $c=1$ (a, b) 
the soliton reflects elastically.  Like above, the boundary 
appears as a hard wall.  However, for wider boundaries, 
$c=4$ (c, d) and $c=10$ (e, f), the soliton dissipates energy through the emission of sound waves.  

It is evident that the reflection of the dark soliton from the soft boundary is typically dissipative (where we are referring to the dissipation of the soliton; the total energy of the system is conserved), although the amount of sound radiated is sensitive to the boundary parameters.  Now we characterise this dissipation in terms of the energy lost from the soliton.  
The  soliton's energy $E_{\rm sol}$  is evaluated numerically before and after the reflection.  This is performed by calculating the energy associated with the soliton within a small region around the soliton, according to the scheme described in Ref. \cite{Parker2010}.  We report the proportional loss in soliton energy after the reflection, normalized with respect to its initial value, and denote this as $\Delta E_{\rm sol}$.

\begin{figure}[htbp]
  \centering
       \includegraphics[width=0.85\columnwidth]{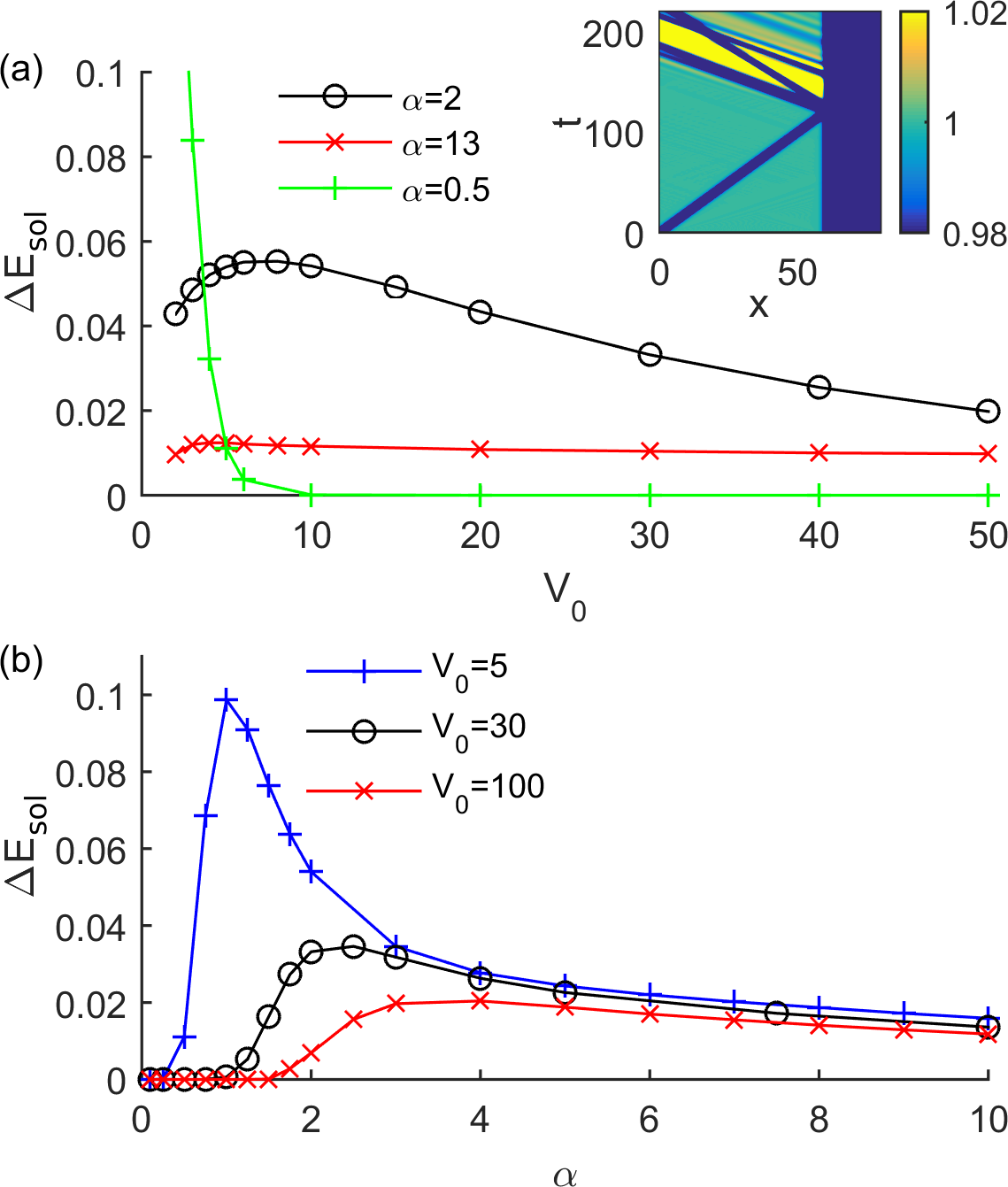}
\caption{(Color online). Energy loss in the dark soliton (normalized by its 
initial energy)  $\Delta E_{\rm sol}$ due to a reflection against 
a power-law boundary. Panel (a) shows  this energy loss as a 
function of the amplitude of the boundary potential $V(L)$, 
for three values of the exponent $\alpha$. Panel (b) shows
$\Delta E_{\rm sol}$ as a function of the exponent $\alpha$ for fixed 
potential amplitude $V_0$.   In (a) the inset is for $\alpha=0.5$ 
and $V_0=2$, showing anomalously high sound emission. 
\label{fig:figure4}}         
\end{figure}

\begin{figure}[htbp]
  \centering
       \includegraphics[width=0.8\columnwidth]{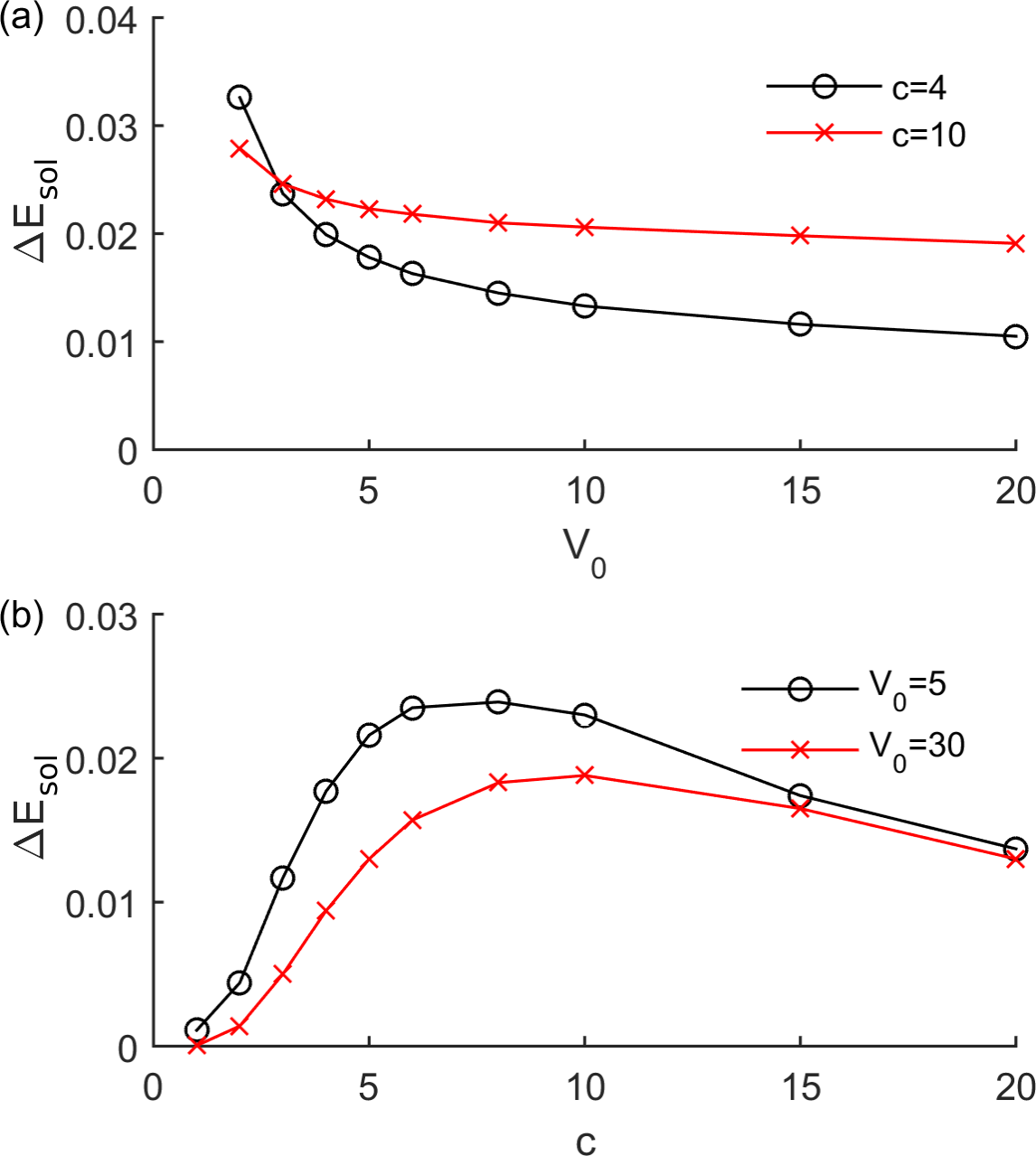}
\caption{(Color online). Energy loss in the dark soliton 
(normalized by its initial energy) $\Delta E_{\rm sol}$
 due to a reflection against an exponential boundary. 
Panel (a) shows this energy loss as a function of the 
amplitude of the boundary potential $V_0$, for two values of the 
Gaussian width $c$, while (b)  displays it 
as a function of the  $c$ for fixed potential amplitude $V_0$.  
\label{fig:figure5}}         
\end{figure}

Figure \ref{fig:figure4}(a) shows the energy loss for the power-law trap as a function of the amplitude of the boundary potential $V_0$, for three values of the potential exponent $\alpha$.  Note that we limit our analysis to $V_0 \geq 2$; below this range the potential does not fully confine the condensate.   For $\alpha=2$ and $\alpha=13$, the energy loss increases to a maximum at moderate $V_0$ ($V_0 \sim 5-10$ for these cases), before decaying with increasinging $V_0$.  This is typical of the general behaviour for $\alpha \geq 1$.  It is worth noting that the softer boundary, $\alpha=2$, gives the most energy loss (up to $5\%$), and that the energy loss decays very slowly with $V_0$, and so causes significant dissipation even for large amplitudes.  For $\alpha<1$, however, the trend is distinct.  For large $V_0$, sound emission is heavily suppressed; this is because for $\alpha < 1$ the boundary potential rises up with a very large gradient (which decreases with distance into the boundary).  As such, for $V_0 \gg 1$ the condensate/soliton effectively experiences a hard wall potential.  For smaller $V_0$, however, the condensate/soliton experiences the low gradient region of the boundary, inducing sound emission.  The energy loss increases rapidly as $V_0$ is decreased towards the value of $2$, enchanced by an unusual effect where sound waves are generated from the boundary even after the soliton has left the boundary (see inset of Fig. \ref{fig:figure4}(b)).

{Figure \ref{fig:figure4}(b) shows the energy loss as a function of the exponent $\alpha$, for three values of the potential amplitude.  The general behaviour is that the energy loss is typically vanishingly small for small $\alpha$, due to the hard-wall effect mentioned above, and is also small for very large $\alpha$, since the potential increases rapidly and also begins to approximate a hard wall.  However, in between these limits, the energy loss reaches maximum; this position of this maximum is dependent on $\alpha$ but typically lies in the range $1<\alpha < 5$.}


{Similarly, we have explored the energy loss from a single reflection of an exponential boundary.  For fixed width $c$ (Fig. \ref{fig:figure5}(a)), the energy loss is highest for the lowest amplitudes, and decreases as $V_0$ is increased.  Meanwhile, for fixed amplitude $V_0$ (Fig. \ref{fig:figure5}(b)) the energy loss is vanishingly small for small width $c$; here the exponential wall is so narrow that it resembles the hard wall.  The energy loss increases with $c$, reaches a maximum for moderate values $c \sim 5-10$, and then decreases slowly with $c$.  The energy loss is typically of the order of a few percent.  }


\subsection{Multiple reflections}\label{sec:multiple}

{In a single reflection, the energy loss from the soliton is small, typically of the order of a few percent, and the increase in its speed is so small that it is not visible by eye.  However, in the course of multiple reflections, such as due to a dark soliton oscillating back and forth in a box trap, significant decay of the soliton can be expected. } 

\begin{figure}[htbp]
  \centering
       \includegraphics[width=0.8\columnwidth]{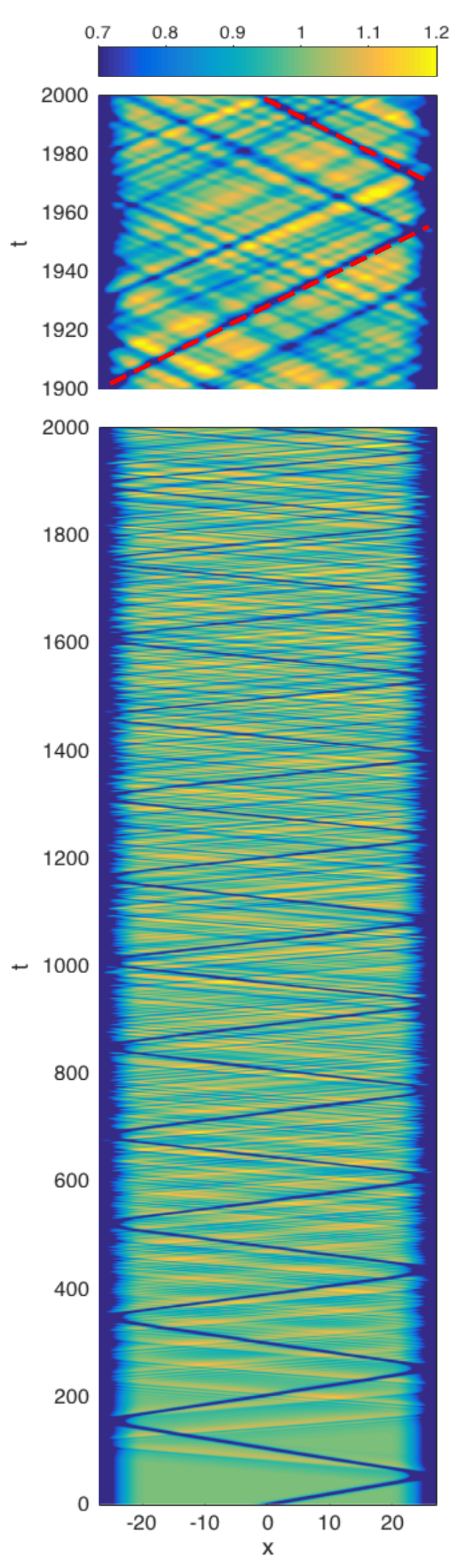}
\caption{{Dark soliton (initial speed $v=0.5$) oscillating in a power-law box trap.  The upper plot shows a zoom-up over the time-range [1900,2000], with the original solition indicated by the dashed red line.   Trap parameters $V_0=5$, $\alpha=2$, $L=40$ and $w=20$.}  \label{fig:figure6}
}         
\end{figure}

{Figure \ref{fig:figure6} shows the long-term evolution of a dark soliton, with initial speed $v=0.5$, oscillating back and forth in a power-law trap (parameters $\alpha=2$, $V_0=5$).  With each reflection the soliton loses amplitude and speeds up, while the condensate becomes increasingly populated with density waves.  After of the order of 25 reflections the soliton has reached a speed $v \sim 0.9$.    Interestingly, at late times (see upper plot), additional fast dark soliton-like structures (low density, localized structures) appear to pass back and forth through the box.  }

 \begin{figure}[htbp]
  \centering
       \includegraphics[width=0.5\textwidth]{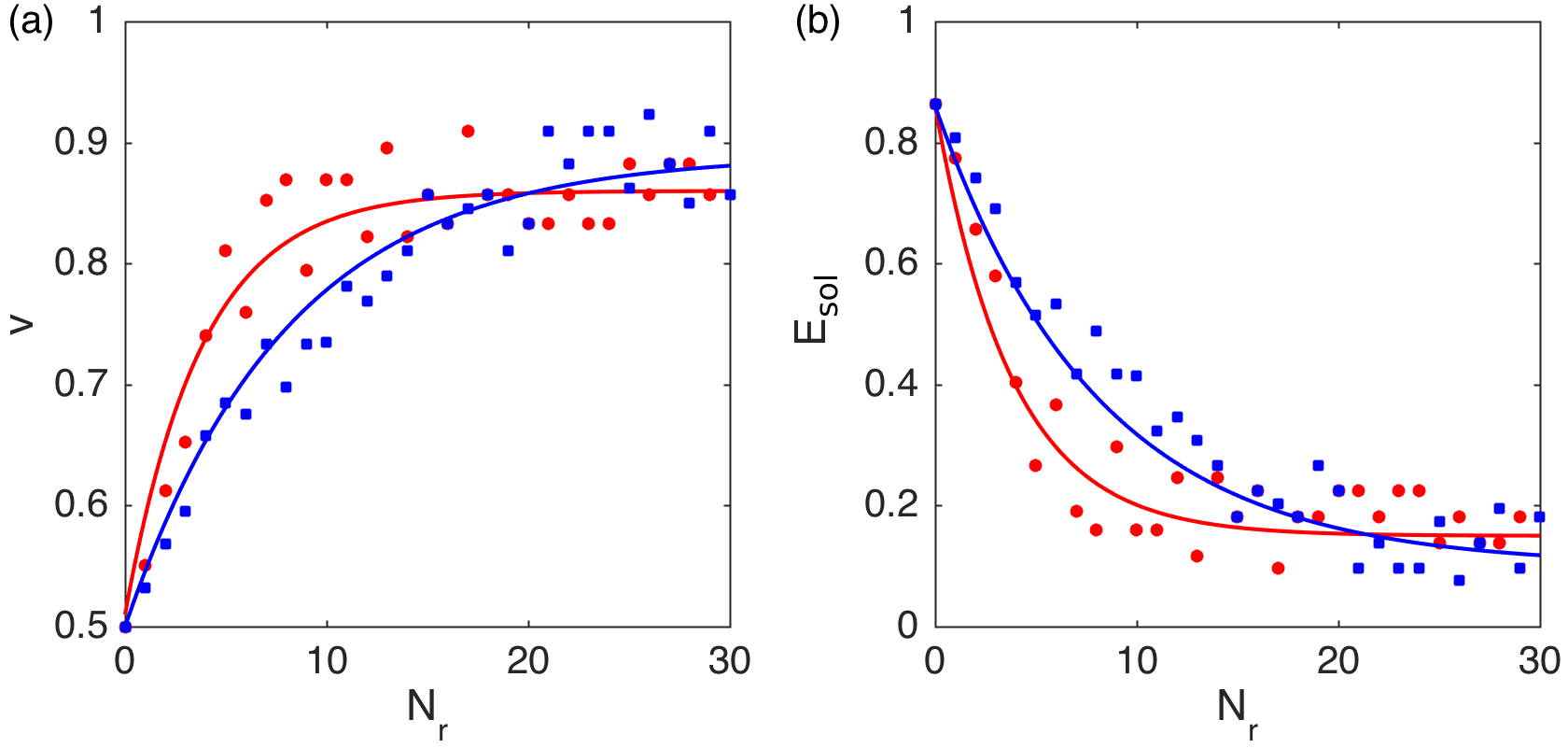}
\caption{{Decay of a soliton with initial speed $v=0.5$ after multiple reflections in a power-law trap.  (a) Soliton speed after $N_{\rm r}$  reflections for $\alpha=1$ (red points) and $\alpha=2$ (blue data).  The speed is measured as the average speed through the bulk of the condensate.  
(b)   The soliton's energy $E_{\rm sol}$,
determined using the speed-energy relation \eqref{eq:energy}.  The solid lines are exponential fits to the data.  Other parameters: $V_0=5$, $L=40$ and $w=20$. }  \label{fig:figure7}}         
\end{figure}

To quantify the decay of the soliton during the repeated oscillations through the box we monitor the speed of the soliton through the bulk of the condensate following each reflection.  Figure \ref{fig:figure7}(a) shows the soliton speed versus the number of reflections $N_{\rm r}$ for two power-law boxes, while Fig. \ref{fig:figure7}(b) shows the corresponding behaviour for 
the soliton's energy, calculated using the  energy-speed relation for a dark soliton \eqref{eq:energy}.  The qualitative behaviour is general: the soliton speed increases, relaxes towards a maximum value (which is less than unity), while the soliton's energy decays towards a value (which exceeds zero).  The trends are captured by an exponential fit (solid lines).  Importantly, these results shows that the soliton does not decay away completely, but saturates towards a high speed/low energy state.  By these late times, the system is full of density waves of similar amplitude, suggesting that the decay may be stabilized by absorption of energy from the density waves.

\section{Discussion}\label{sec:discussion}

We have seen that the reflection of the dark soliton from a soft wall is typically dissipative, in that the soliton loses energy through the emission of sound waves.  An explanation of the partial reflection of the soliton can be found in the context  of the propagation of optical pulses in an optical fiber.  Indeed, the evolution equation of a dark solition in a normal-dispersion optical fiber is the so-called nonlinear Schr\"odinger equation, which is essentially the Gross-Pitaevskii equation\;\eqref{eq:GP}  with  time and spatial coordinates inverted.  It is known there that the propagation of dark soliton  is guaranteed thanks to the balancing between the linear term $u_{xx}$ (dispersion relation) and the nonlinear term $|u|^2 u$ (self-phase modulation), as it occurs in our experiments where $V(x)=0.$  When the soliton approaches the potential wall, the balancing between the linear term $u_{xx}$ and $|u|^2 u$ no longer occurs since the latter term now becomes $\left(|u|^2+V(x)\right) u$.  In the optical context, the presence of $V(x)$ corresponds to a modification of the refractive index, which is equivalent to say that some energy of the soliton is supplied to generate some harmonic waves which travel faster than all the harmonic waves making their  ``envelope", that is, the dark soliton wave.

\begin{figure}[htbp]
  \centering
       \includegraphics[width=0.4\textwidth]{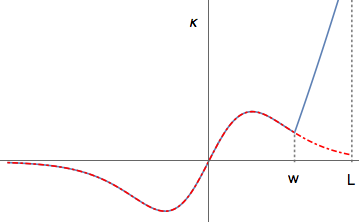}
\caption{The instantaneous wavenumber $\kappa$ versus the $x$ axis for $\alpha=2$. The dot-dashed red line refers to $V(x)=0$, whereas the blue one to $V(x) \sim x^{2}$. The black solid line focus the center of the dark soliton and the dashed grey lines, instead, mark  the region where the potential takes place.  }  \label{fig:figure8}         
\end{figure}

\begin{figure}[htbp]
  \centering
       \includegraphics[width=0.4\textwidth]{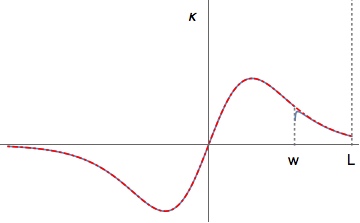}
\caption{The same configuration used in Fig.\;\ref{fig:figure8} with  $\alpha=0.001$. }  \label{fig:figure9}         
\end{figure}

To make more evident how the potential $V(x)$ modifies the mechanism of ``spectral broadening'' of the dark soliton caused by nonlinearity, we consider equation\;\eqref{eq:GP} without the dispersion term $u_{xx}$ (which is  responsible only for the dispersion mechanism).  
The wave solution  is then    straightforward to find, and it takes the form $u(x,t)=|u(x,0)| \exp(\j \phi )$ with the phase $\phi=(|u(x,0)|^2+V(x))t$ depending also on $x$, which implies that the instantaneous wavenumber $\kappa$  differs across the wave from the central value. In Figs.\;\ref{fig:figure8}  and  \ref{fig:figure9}  we show  how the wavenumbers $\kappa$ of the dark soliton are  shifted by the nonlinearity and the potential wall by plotting $\partial \phi/\partial x$ for $\alpha=2$ and $\alpha=0.001$   when the dark soliton approaches the potential wall. The dot-dashed red line refers to the absence of potential (which occurs in the central region of the BEC), while the blue line to the presence of the potential wall in the region between $w$ and $L$. Note that for $\alpha=0.001$ (Fig.\;\ref{fig:figure9}) the two curves perflectly match, namely the potential wall does not affect the dark soliton which keeps on running undisturbed. However, for $\alpha=2$ (Fig.\;\ref{fig:figure8})  the potential strongly modifies the instantaneous  wavenumber $\kappa$, shifting it upwards and causing a different distribution of the energy (energy is supplied to other harmonics which run away from the soliton). 
 
As shown in Figures  \ref{fig:figure2}, \ref{fig:figure3} and \ref{fig:figure6}, after the interaction of the soliton with the side of the potential, the ``new" dark soliton (with lower energy and higher speed) enters again in the region with null potential, where the GP is completely integrable. The important issue here is that GP \eqref{eq:GP} admits the  solution \eqref{sol:GP} for any $k$, namely the ``new" dark soliton (just recovered from the side) may propagate undisturbed again in the BEC without any loss and showing its main features, as for instance to keep its identity after a collision with the other waves (in our case, the sound waves). 

Over multiple oscillations in the box-like trap, the energy loss and speed increase of the soliton (which is very small for a single reflection) can become significant. With each reflection the condensate becomes increasingly populated with dispersive density waves, which are soon well-distributed through the condensate.  This procedure lasts untill the density depth $k$ of the soliton is comparable to the amplitude of these dispersive waves (see Fig.\;\ref{fig:figure6}).  It is then hard to distinguish the residual dark soliton from  the overlapping waves  (see the top of Fig.\;\ref{fig:figure6}), causing two complications.  Firstly, the evaluation of the speed or energy of the soliton becomes affected by these waves overlapping the soliton (causing the scatter in the points in Fig. \ref{fig:figure7}).  Secondly, the interaction of the soliton  with these dispersive waves cannot be neglected, even though the soliton keeps its identity after the reflections \cite{Oreshnikov-OL2015}. Indeed, density waves (sound) can supply energy back to some harmonics of the soliton, enhancing its energy.

\section{Conclusions}\label{sec:conclusions}

We have studied the propagation of dark solitons in
1D zero-temperature Bose-Einstein condensates confined by box-like external 
potentials consisting of
a central flat region where the condensate is pratically free
 and two ``soft" walls (power-law or Gaussian) of variable 
height and steepness.  In the central region
the one-dimensional GP is completely integrable,
and dark solitons propagate undisturbed. 
When a soliton meets a side of the trap,
depending on the steepness, the soliton may experience
total or partial reflection. In the case of partial reflection, small
amplitude  density waves (sound) are generated and carry energy away 
from
the soliton, and the soliton's speed increases slightly.  
 We map this 
energy loss as a function of the wall parameters. The reflection is 
perfect for almost vertical sides.  In the dissipative regime and for 
multiple reflections, the soliton's decay
becomes significant.  
The condensate becomes increasingly populated by dispersive density 
waves and when the soliton's depth  reaches the level of 
these waves, its  decay stabilizes.   Finally, we can conclude that the stability and dynamics of dark solitons in box-like traps is fundamentally distinct from that in the well-studied case of harmonic potentials, where the soliton is established to propagate with no net dissipation.

\section*{Acknowledgements}
M.S.   acknowledge the financial support of the Istituto Nazionale di Alta Matematica (GNFM--Gruppo Nazionale della Fisica Matematica).    N. G. P. acknowledges funding from the Engineering and Physical Sciences Research Council (Grant No. EP/M005127/1).


\begin{thebibliography}{10}

\bibitem{Kivshar1998} Y. S. Kivshar and B. Luther-Davies, {\it Phys. Rep.} {\bf 278}, 81 (1998).

\bibitem{Emplit}
P.~Emplit, J.P. Hamaide, F.~Reynaud, C.~Froehly, and A.~Barthelemy, {\em Optics Comm.} {\bf 62}, 374 (1987).

\bibitem{Krokel}
D.~Kr\"okel, N.~J. Halas, G.~Giuliani, and D.~Grischkowsky, {\em Phys. Rev. Lett.} {\bf 60}, 29 (1988).

\bibitem{Weiner}
A.~M. Weiner, J.~P. Heritage, R.~J. Hawkins, R.~N. Thurston, E.~M. Kirschner,
  D.~E. Leaird, and W.~J. Tomlinson, {\em Phys. Rev. Lett.} {\bf 61}, 2445 (1988).
  
  \bibitem{Chen} M. Chen, M. A. Tsankov, J. M. Nash and C. E. Patton, 
{\it Phys. Rev. Lett.} {\bf 70}, 1707 (1993)
  
  \bibitem{Heidemann}
R.~Heidemann, S.~Zhdanov, R.~S\"utterlin, H.~M. Thomas, and G.~E. Morfill, 
{\it Phys. Rev. Lett.}, {\bf 102}, 135002 (2009).

\bibitem{Smirnov}
E. Smirnov, C.~E. R\"uter, M. Stepi\ifmmode~\acute{c}\else
  \'{c}\fi{}, D. Kip, and V. Shandarov, {\em Phys. Rev. E} {\bf 74}, 065601 (2006).
  
  \bibitem{Chabchoub}
A.~Chabchoub, O.~Kimmoun, H.~Branger, N.~Hoffmann, D.~Proment, M.~Onorato, and
  N.~Akhmediev, {\em Phys. Rev. Lett.} {\bf 110}, 124101 (2013).
  

  
  \bibitem{Frantzeskakis2010}
D~J Frantzeskakis, {\em J. Phys. A} {\bf 43}, 213001  (2010).

\bibitem{Stenholm-PR363-2002heuristic}
S. Stenholm, {\em Phys. Rep.} {\bf 363}, 173 (2002).

\bibitem{Carretero-N21-2008nonlinear}
R.~Carretero-Gonz{\'a}lez, D. J.~Frantzeskakis, and P. G.~Kevrekidis,
{\em Nonlinearity} {\bf 21}, R139 (2008).

\bibitem{Gross-NC20-1961structure}
E.~P. Gross, {\em Il Nuovo Cimento (1955-1965)} {\bf 20}, 454 (1961).

\bibitem{Gross-1963}
E.~P. Gross, {\em J. Math. Phys.} {\bf 4}, 195 (1963).

\bibitem{Pitaevskii-1961}
LP~Pitaevskii, {\em J. Exp. Theor. Phys.} {\bf 13}, 451  (1961).

\bibitem{Burger1999}
S. Burger, K. Bongs, S. Dettmer, W. Ertmer, K. Sengstock, A. Sanpera, G. V. Shlyapnikov, and M. Lewenstein, {\it Phys. Rev. Lett.} {\bf 83}, 5198 (1999).

\bibitem{Denschlag2000}
J. Denschlag, J. E. Simsarian, D. L. Feder, C. W. Clark, L. A. Collins, J. Cubizolles, L. Deng, E. W. Hagley, K. Helmerson, W. P. Reinhardt, S. L. Rolston, B. I. Schneider, W. D. Phillips, {\it Science} {\bf 287}, 97 (2000).

\bibitem{Dutton2001} Z. Dutton, M. Budde, C. Slowe and L. V. Hau, {\it Science} {\bf 293}, 663 (2001).


\bibitem{Jo2007} G.-B. Jo, J.-H. Choi, C. A. Christensen, T. A. Pasquini, Y.-R. Lee, W. Ketterle, and D. E. Pritchard, {\it Phys. Rev. Lett.} {\bf 98}, 180401 (2007).

\bibitem{Engels2007} P. Engels and C. Atherton, {\it Phys. Rev. Lett.} {\bf 99}, 160405 (2007).

\bibitem{Becker2008}
C. Becker, S. Stellmer, P. S.-Panahi, S. D\"orscher, M. Baumert, E.-M. Richter, J. Kronj\"ager, K. Bongs, and K. Sengstock, {\it Nat. Phys.} {\bf 4}, 496 (2008). 

\bibitem{Chang2008} J. J. Chang, P. Engels and M. A. Hoefer, {\it Phys. Rev. Lett.} {\bf 101}, 170404 (2008).

\bibitem{Stellmer2008}
S. Stellmer, C. Becker, P. Soltan-Panahi, E.-M. Richter, S. D\"orscher, M. Baumert, J. Kronj\"ager, K. Bongs, and K. Sengstock, {\it Phys. Rev. Lett.} {\bf 101}, 120406 (2008).

\bibitem{Weller2008}
A. Weller, J. P. Ronzheimer, C. Gross, J. Esteve, M. K. Oberthaler, D. J. Frantzeskakis, G. Theocharis, and P. G. Kevrekidis, {\it Phys. Rev. Lett.} {\bf 101}, 130401 (2008).

\bibitem{Aycock2016} L. M. Aycock, H. M. Hurst, D. Genkina, H.-I. Lu, V. Galitski and I. B. Spielman, arXiv:1608.03916 (2016).

  \bibitem{Reinhardt1997} W. P. Reinhardt and C. W. Clark, {\it J. Phys. B} {\bf 30}, L785 (1997).
  
\bibitem{Busch2000} Th. Busch and J. R. Anglin, {\it Phys. Rev. Lett.} {\bf 84}, 2298 (2000).

\bibitem{Huang2002} G. Huang, J. Szeftel and S. Zhu, {\it Phys. Rev. A} {\bf 65}, 053605 (2002).


  \bibitem{Parker2003} N. G. Parker, N. P. Proukakis, M. Leadbeater and C. S. Adams, {\it Phys. Rev. Lett.} {\bf 90}, 220401 (2003).
  
  \bibitem{Pelinovsky2005} D. E. Pelinovsky, D. J. Frantzeskakis and P. G. Kevrekidis, {\it Phys. Rev.} E {\bf 72}, 016615 (2005)
  
  \bibitem{Parker2010} N. G. Parker, N. P. Proukakis and C. S. Adams, 
{\it Phys. Rev. A} {\bf 81}, 033606 (2010).
  
  \bibitem{Pelinovsky1996} D. E. Pelinovsky, Y. S. Kivshar and V. V. Afanasjev,
{\it Phys. Rev. E} {\bf 54}, 2015 (1996).


  
  \bibitem{Fedichev1999} P. O. Fedichev, A. E. Muryshev, and G. V. Shlyapnikov, {\it Phys. Rev. A} {\bf 60}, 3220 (1999).
  
  \bibitem{Muryshev2002} A. E. Muryshev, G. V. Shlyapnikov, W. Ertmer, K. Sengstock and M. Lewenstein, {\it Phys. Rev. Lett.} {\bf 89}, 110401 (2002).
  
  \bibitem{Jackson2007} B. Jackson, N. P. Proukakis and C. F. Barenghi, 
{\it Phys. Rev. A} {\bf 75}, 051601(R) (2007).

  \bibitem{Theocharis2007} G. Theocharis, P. G. Kevrekidis, M. K. Oberthaler and D. J. Frantzeskakis, {\it Phys. Rev. A} {\bf 76}, 045601 (2007).


\bibitem{Muryshev1999} A. E. Muryshev, H. B. van Linden van den Heuvell, and G. V. Shlyapnikov, {\it Phys. Rev. A} {\bf 60}, R2665 (1999).

\bibitem{Carr2000} L. D. Carr, M. A. Leung and W. P. Reinhardt, {\it J. Phys. B: At. Mol. Opt. Phys.} {\bf 33}, 3983 (2000).

\bibitem{Feder2000} D. L. Feder, M. S. Pindzola, L. A. Collins, B. I. Schneider and C. W. Clark, {\it Phys. Rev. A} {\bf 62}, 053606 (2000).

\bibitem{Anderson2001} B. P. Anderson {\it et al.}, 
{\it Phys. Rev. Lett.} {\bf 86}, 2926 (2001).

\bibitem{Hoefer2016} M. A. Hoefer and B. Ilan, {\it Phys. Rev. A} {\bf 94}, 013609 (2016).


\bibitem{Parker2003b} N. G. Parker, N. P. Proukakis, M. Leadbeater 
and C. S. Adams, {\it J. Phys. B} {\bf 36}, 2891 (2003).

\bibitem{Allen2011}  A. J. Allen, D. P. Jackson, C. F. Barenghi and N. P. Proukakis, {\it Phys. Rev. A} {\bf 83}, 013613 (2011).


\bibitem{Kevrekidis2003} P. G. Kevrekidis, R. Carretero-Gonzalez, G. Theocharis,
D. J. Frantzeskakis, and B. A. Malomed, {\it Phys. Rev. A}
{\bf 68}, 035602 (2003).

\bibitem{Parker2004} N. G. Parker {\it et al}. 
{\it J. Phys. B} {\bf 37}, S175 (2004).

\bibitem{Frantzeskakis2002} D. J. Frantzeskakis {\it et al.},  {\it Phys. Rev. A} {\bf 66}, 053608 (2002).

\bibitem{Proukakis2004} N. P. Proukakis, N. G. Parker, D. J. Frantzeskakis 
and C. S. Adams, {\it J. Opt. B} {\bf 6}, S380 (2004).

\bibitem{Radouani2003} A. Radouani, {\it Phys. Rev. A} {\bf 68}, 043620 (2003).

\bibitem{Radouani2004} A. Radouani, {\it Phys. Rev. A} {\bf 70}, 013602 (2004).

\bibitem{Bilas2005a} N. Bilas and N. Pavloff, {\it Phys. Rev. A} {\bf 72}, 033618 (2005).

\bibitem{Hans2015} I. Hans, J. Stockhofe, and P. Schmelcher, 
{\it Phys. Rev. A} {\bf 92}, 013627 (2015).

\bibitem{Bilas2005b} N. Bilas and N. Pavloff, 
{\it Phys. Rev. Lett.} {\bf 95}, 130403 (2005)

%



  

 \bibitem{Meyrath} 
 T. P. Meyrath, F. Schreck, J. L. Hanssen, C.-S. Chuu, M. G. Raizen, 
{\it Phys. Rev. A} {\bf 71}, 041604(R) (2005).
 
 \bibitem{VanEs} J. J. P. van Es, P. Wicke, A. H. van Amerongen, C. R\'etif, S. Whitlock and N. J. van Druten, {\it J. Phys. B} {\bf 43}, 155002 (2010)
 
 \bibitem{Chomaz} L. Chomaz, L. Corman, T. Bienaime, R. Desbuquois, 
C. Weitenberg, S. Nascimbene, J. Beugnon and J. Dalibard, 
{\it Nat. Comm.} {\bf 6}, 6162 (2015).
 
 \bibitem{Gaunt} 
A.L. Gaunt, T.F. Schmidutz, I. Gotlibovych, R.P. Smith and Z. Hadzibabic, 
{\it Phys. Rev. Lett.} {\bf 110}, 200406 (2013).




\bibitem{BruguarinoJMP51-2010}
T.~Brugarino and M.~Sciacca,
 {\em J. Math. Phys.} {\bf 51}, 093503 (2010).
 
\bibitem{Primer} C. F. Barenghi and N. G. Parker, ``A Primer on Quantum Fluids" (Springer, Berlin, 2016)


%
%
%
%
%
%
%
%
%
%
%
%
%


\bibitem{Oreshnikov-OL2015}
I.~Oreshnikov, R.~Driben, and A.V.~Yulin.  {\em Opt. Lett.} {\bf 40}, 4871 (2015).


\end{thebibliography}
\end{document}